# Digital Surveillance Networks of 2014 Ebola Epidemics and Lessons for COVID-19

**Liaquat Hossain, Fiona Kong, and Derek Kham**

Nebraska Healthcare Collaborative Chair of Population Health, Department Chairperson, Department of Cyber Systems, College of Business and Technology, The University of Nebraska at Kearney, NE 68849, Email: hossainl2@unk.edu

Research Associate, City University of Hong Kong, Hong Kong

Senior Research Assoc, Division of Information & Tech Studies, University of Hong Kong, HK

**Abstract**
What lessons can be learned from the management of the 2014 Ebola outbreaks so that COVOID-19 and the ongoing variant surveillance? In this paper, we argue that effective management of outbreaks, like the West African 2014 Ebola epidemic, is dependent on the use of multi method approach to detect public health preparedness. We are increasingly seeing a delay and disconnect of the transmission of locally situated information to the hierarchical system for making the overall preparedness and response more proactive than reactive for dealing with complex emergencies such as 2014 Ebola. For our COVID-19, we also observed institutional and public behaviour similar to 2014 Ebola response. It is timely to consider whether digital surveillance networks and support systems can be used to bring the formal and community based ad hoc networks required for facilitating the transmission of both strong (i.e., infections, confirmed cases, deaths in hospital or clinic settings) and weak alters from the community. This will allow timely detection of symptoms of isolated suspected cases for making the overall surveillance and intervention strategy far more effective. The use of digital surveillance networks can further contribute to the development of global awareness of complex emergencies such as Ebola for constructing information infrastructure required to develop, monitor and analysis of community based global emergency surveillance in developed and developing countries. In this study, a systematic analysis of the spread during the months of March to October 2014 was performed using data from the Program for Monitoring Emerging Diseases (ProMED) and the Factiva database. Using digital surveillance networks, we aim to draw network connections of individuals/groups from a localized to a globalized transmission of Ebola using reported suspected/probable/confirmed cases at different locations around the world. We argue that public health preparedness and response can be strengthened by understanding the social network connections between responders (such as local health authorities) and spreaders (infected individuals and groups).

Key words: Digital Surveillance, Social Networks, Ebola, Epidemics

**Introduction**

Ebola virus disease (EVD) or Ebola haemorrhagic fever is a disease which can affect humans and/or other primates[1, 2]. It is caused by ebolaviruses, which belong to the viron family of Filoviridae. The current outbreak strain is the Zaire Ebola virus (EBOV, formerly ZEBOV)[3]. There are four other known species of Ebola; Bundibugyo ebolavirus, Reston ebolavirus (REBOV), Sudan ebolavirus (SEBOV), and Taï Forest ebolavirus (originally Côte d'Ivoire ebolavirus)[3, 4]. Of the five viruses, only REBOV is not known to cause disease in humans[2]. The first reported cases of Ebola occurred in two outbreaks which happened simultaneously in Nzara, Sudan, and Yambuku, Democratic Republic of Congo (DRC) [4, 5]. Ebola took its name after the Ebola River near one of the affected villages. The Sudanese strain, SEBOV, had a mortality rate of 53% (150/284) and the DRC strain (EBOV) had a mortality rate of 89% (284/318). In the years of 1977 and 1979, two more Ebola outbreaks occurred in Tandala, Zaire [6] and South Sudan respectively [7]. After a hiatus of 15 years, it re-emerged in Gabon with isolated types related to the DRC strain in 1994 [8]. In 1995, an Ebola outbreak started in Kikwit, DRC with an estimated mortality rate of 81 per cent. From 2000 to 2012, there were waves of Ebola outbreaks reported in Africa with a break between 2004 and 2007.

The incubation period for Ebola virus in humans is between two to twenty one days [2, 9]. Signs and symptoms typically consist of fever, sore throat, muscle pain and headaches. The disease progresses to vomiting, diarrhea and rashes then followed by poor liver and/or kidney function [9, 10]. Internal and external haemorrhaging may occur in some individuals[9]. The death rates vary differently between strains but the case-fatality rate can go as high as 90 per cent [2, 9, 11, 12]. Death is usually caused by fluid loss and follows between six to sixteen days after the symptoms appear. The virus spreads mainly by direct contact with bodily fluids (e.g., blood, semen, breast milk) of an infected human or animal[13]. Direct contact with a recently contaminated item or surface is also a factor in the transmission of the virus[9, 14]. There are no known cases of airborne transmission. The natural carriers of Ebola are African fruit bats which are not affected by it [2]. Humans can become infected by contact with a carrier bat or a living or dead animal which has been infected by the bat. The main issue with identifying Ebola is that its symptoms mimicked other common infectious diseases in the region. These diseases include malaria, cholera, typhoid fever, and other viral haemorrhagic fevers [9, 11, 15]. Only a blood sample tested for the viral RNA can confirm the presence of Ebola.

The Program for Monitoring Emerging Diseases (ProMED), an Internet-based reporting system dedicated to rapid global dissemination of information on disease outbreaks in humans and animals, was alerted to a report in Standard Media Kenya. The report referred to a localized outbreak of unknown viral haemorrhagic fever which had occurred in the border village of Guéckédou Prefecture, Guinea[16, 17]. Within a few months, it had reached epidemic status and affected the neighboring countries of Liberia and Sierra Leone[17]. Small isolated outbreaks have been known to occur in sub-Saharan Africa but no cases have ever been reported in Guinea [5, 6, 8, 11]. Over the past few months from March to October 2014, a ProMED request for information (RFI) on an isolated outbreak in a border Guinean village turned into the West African Ebola epidemic with exported cases to other regions of the world. By then, the outbreak magnified and propagated to larger populations around the African region and an exported case in USA. This led to the declaration of an International Health Emergency by the World Health Organization (WHO) under the International Health Regulations (2005) [IHR (2005)] on 7 August 2014 [18].

The WHO Ebola Response Roadmap Situation Report of 15 October 2014 highlights that there are 8,997 confirmed, probable, and suspected cases of Ebola virus disease (EVD) which have been reported in seven affected countries (Guinea, Liberia, Nigeria, Senegal, Sierra Leone, Spain, and the United States of America) causing 4,493 deaths (almost 50% mortality rate)[19]. This spread has also affected the health-care workforce with 427 confirmed cases and 236 deaths (55% mortality rate). We are further seeing evidences that a number of countries such as Nigeria, Senegal, Spain and the US with localized transmission with reported cases imported from a country with wider and intense transmission [19]. The U.S. Centers for Disease Control and Prevention (CDC has predicted that Ebola cases will amount to 20,000 per week by December 2014. Moreover, the CDC released a report on September 16, 2014 predicting as many as 550,000 to 1.4 million cases of the Ebola virus in Liberia and Sierra Leone alone, by January 20, 2015, according to two worst-case scenarios from scientists studying the historic outbreak [20].

Control of outbreaks often requires coordinated medical services with community engagement. Rapid assessment, contact tracing, isolation of infected individuals, proper cremation/burial of the dead and access to laboratory services are necessary to the containment of Ebola [13, 14, 21, 22]. Preventative measures include community education on proper handling of potentially infected bush or bat meat while wearing protective clothing and thorough cooking[9]. The West African 2014 Ebola epidemic demonstrated the highly interconnected and interdependent social systems in the region and the rest of the world. Failure to contain the disease and the lack of effective responses at the hierarchical level from the bottom to the top (i.e., villages → towns → suburbs → counties/province → country) could have significant and devastating impact on the global population. A wider recognition and a stronger need to move towards digital surveillance networks which could bring two systems—hierarchical or formalized command control driven approach and community based or ad hoc emerging networks is necessary to fill the missing gaps and links to make the whole system more responsive and effective in a proactive manner than the current reactive strategies that we have seen for dealing with the Ebola spread.

**Digital Surveillance Networks of Epidemics**
Digital disease surveillance is a recent advancement in early detection of infectious diseases and/or research of other diseases (e.g. mood disorders). This form of surveillance have branched out from internet based technologies in the earlier years [23]to mobile technologies.[24] Over the last two decades, current developments have gone beyond human health with applications extending to even animal health and monitoring of vector-borne diseases, transport modelling, as well as spatiotemporal mapping of unusual health events harnessing information on the internet. [25, 26, 27] Digital disease surveillance is based on the extraction of data through formal (i.e. government controlled and sanctioned information) and informal electronic information to identify abnormal/unusual events which may occur in the community or a given population in a geographical locality during a specified period. Examples of internet based usage are mostly informal with the use of search query data,[28] online news, [29, 30] and social media/networks [23, 30, 31] while mobile technologies include formal data such, like call detail records (CDRs), which identify the location of the original call made, and even smart phones with global positioning systems (GPS).[24, 25] The concept of online disease surveillance has recently experienced significant realizations and progresses. Various online syndromic surveillance methods and models are being continuously developed, assessed and improved, using data from i) search engine queries (such as Google queries), ii) social network blogs (*e.g.* Twitter microblogs) or iii) health website accesses (*e.g.* Wikipedia visits). With the first of these studies appearing in 2004 [32], the past 10 years have seen another close to 70 studies in the area [33; Fig.1]. There are roughly equal proportions of Google search queries (Google Flu Trends and Google Trends) and Twitter microblogs studies. Notice that several studies use Google Trends data, which are available for more regions and provide more flexibility for modeling than the ready-made Google Flu Trends.

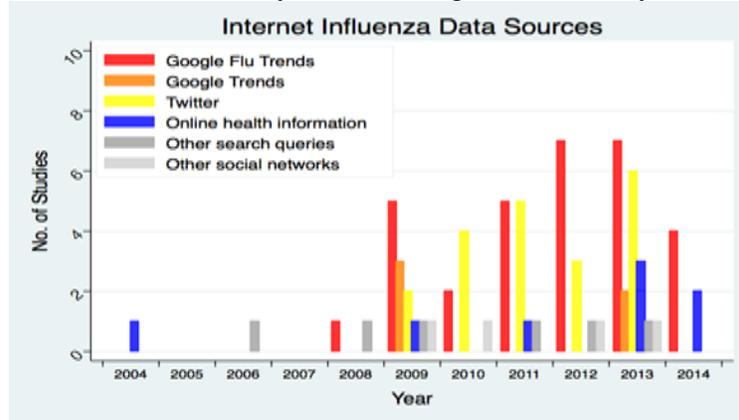

Figure 1. Distribution of Internet data sources of 65 online influenza syndromic surveillance

The studies have been reporting accurate influenza estimations (Figure 2), with each kind of data conferring specific advantages for the studies. Google Flu Trends (www.google.org/flutrends) reported a very high correlation of 0.94 [34] with official US CDC sentinel physician influenza-like-illness (ILI) data. Google has kept the algorithm confidential and discontinued estimations in August 2015, passing the data on to research organizations such as Healthmap (www.healthmap.org). Studies using Twitter microblogs and Wikipedia access logs as their data are also reporting correlations around the 0.9 level. Whereas search engine query studies include an almost unprecedented abundance of data, social network blogs provide more contextual information (such as in terms of semantics and social relationships) for data interpretation and network modeling. Whereas both Google Flu Trends and Google Trends algorithms and data remain behind veil, Wikipedia data are freely available [35, 36] and relatively standardized across languages and diseases [35].

Informal data from social media can also include participatory crowdsourcing either through citizens or professional groups,[23, 26] Twitter monitoring[25, 31] and even Google trends[25]. One example of participatory crowd sourcing was the call for online volunteers by Humanitarian OpenStreetMap Team (HOT) during the 2014 West Africa Ebola crisis to map the cities in Guéckédou, Guinea, based on satellite information of the area and was proven critical for canvassing inhabitants and mapping the Ebola transmission in the city.[25] Other sources of data which have proven useful in the Ebola crisis was HealthMap which identified news stories prior to the official notification of the 2014 West Africa Ebola outbreak.[26] Despite the poor international response to the Ebola outbreak, the use of big data outlined the advantage of it as a complementary system to flag events and identify increasing trends which may warrant further investigation rather than the use as a primary surveillance because of possible internet generated 'noise' and even analysis by poorly trained/untrained public health staff.[23, 25, 26, 27, 30, 37] Integrating several data sources (even social media data) within a digital surveillance system may prove complicated as public health personnel need to be trained prepared and regularly updated or the whole system could fail. [37]

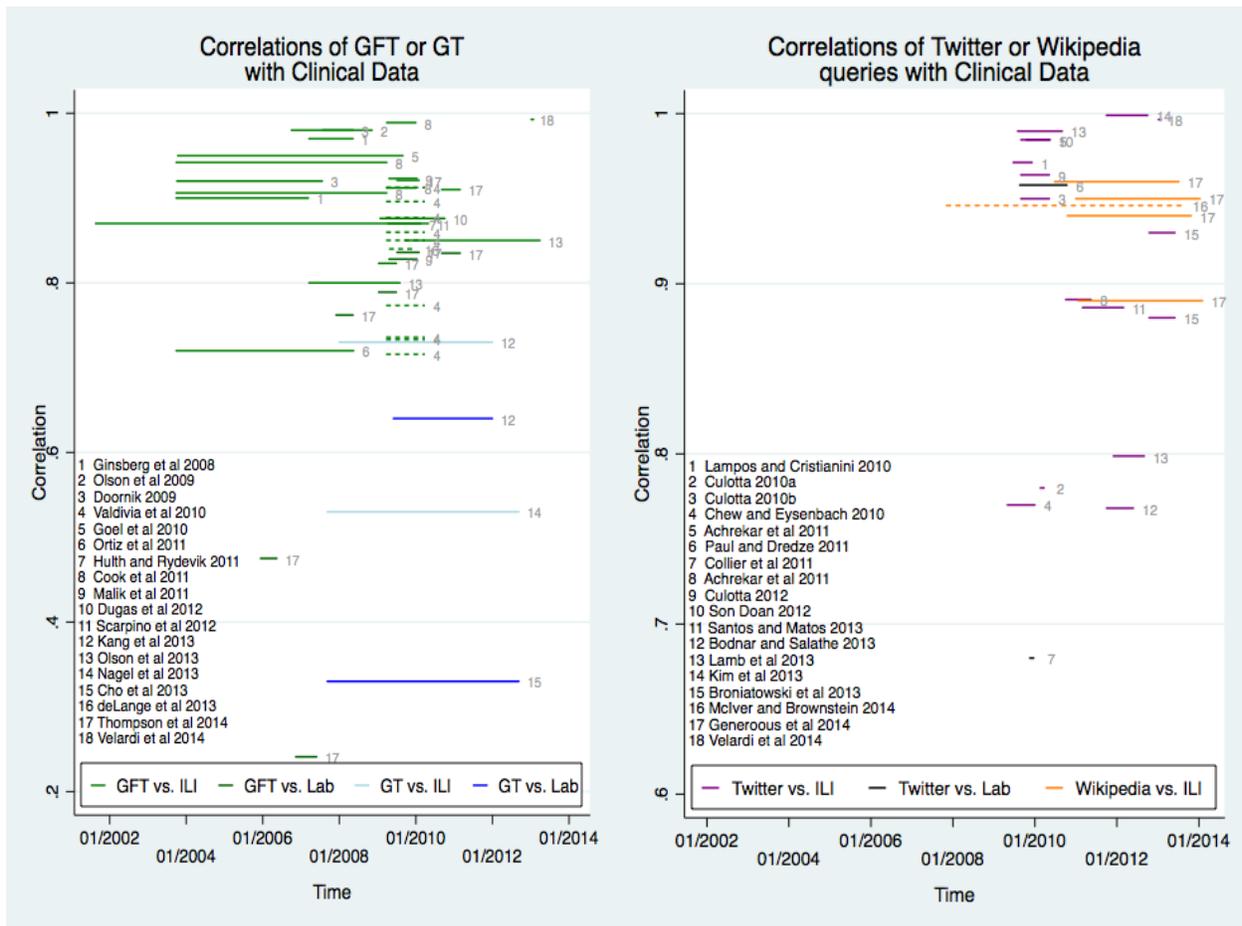

Figure 2. Studies reporting accurate influenza estimations

In the case of the 2014 West African Ebola crisis, the infrastructure of internet and mobile based technologies in the affected countries are not as developed or advanced when compared to developed countries, especially in the rural and regional areas. Mobile phones and the internet usage remains a costly luxury for the rural and regional populations. While monitoring social media in the rural and regional communities may not be feasible in terms of cost, integrating location data on mobile phone users (if the population is large enough) together with surveillance of online news reporting by neighboring or local media may still prove useful in identifying outbreaks of concern. Hence the remaining feasible and cost-efficient option is the combined use of news media and other information based media, with other mobile technologies and surveys, to identify potential hotspots for future outbreaks. However with the right resources and training, this may assist low-resourced developing settings in the development of contingency plans to contain an outbreak and a low-cost complementary surveillance system.

The concept of social network has been successfully applied in various classical epidemiological studies, such as epidemiology, infometrics, sociology, computation theories and disaster medicine and public health preparedness. In the early 1990s, the Save our Sisters Project in North Carolina trained lay health advisors to reach out to their social networks, in order to improve breast cancer awareness and screening seeking behavior[38]. In a related study, Kang *et al.* quantified social ties and found association with increased use of mammography[39]. In the Colorado Springs Study, which ran from 1985 to 1999, the identified cyclic and dendritic social network structures were used to explain the low to moderate level of HIV propagation amongst university students [40]. Using data from the Framingham Heart Study (1971 to present), Christakis, Fowler and Rosenquist have shown that social network plays a major explanatory role in patterns of obesity[41], happiness [42]and depression [43]in the population.

Compared with traditional epidemiological methods, social network models may enrich the analysis in a number of ways. First, patterns of disease transmission can be identified through visualization of the network, where there are clusters of cases and propagation through time. Second, etiological explanations in terms of physical causes and risk factors can be brought into context with the human agents, taking into account various social, interactive effects. Third, response planning and intervention strategies can be systematically analyzed, simulated and tested, both at the hierarchical, command-and-control level, and at the community-wise, *ad hoc* level. The spread and control of the present Ebola epidemic can be readily analyzed by a social network by exploring Ebola cases identified in the media which can assist in connecting different public available data with each other, such as in terms of geographical propinquity, individual's social relationship, and the probable cause of infection, visualized over time. Parameters specific to "contact networks"—networks modeling the spread of infectious diseases involving person-to-person contacts—are included: the direction of transmission, individuals' health statuses, relationships and occupations[44]. Background information such as demographic and lifestyle may also be inputted into the model, although details on individual cases can at time be limited.

**Methods**
Data sources were systematically collected from media reports, the WHO and the International SOS. Due to Ebola, declared as an International Health Emergency, International news agencies such as the Cable News Network (CNN), British Broadcasting Corporation (BBC), Reuters, Wall Street Journal and the Voice of America (VOA) were used to extract details of the location where the cases are confirmed with its links to original destinations highlighting details of known individuals carrying the virus. Some local news agencies, like American based ABC News, The Washington Post; Nigeria based Vanguard; Germany based Deutsche Welle; Norwegian based The Norway Post; Liberia based Daily Observer also provided additional information about Ebola cases. News and information gathering platforms, like Google News, Yahoo News, Wikipedia, provided further information for indexing. Situational reports from the WHO were used to extract official data and cases details for Ebola confirmed cases and deaths. When possible, individual cases were back tracked from all media sources using Google in order to get a full picture of Ebola through the media and official data. An analysis of the media attention will also be reported. All data were compared and merged.

As an added supplement, a search was performed in ProMED with the term 'Ebola' in the two categories of Post and Subject keywords (see figure 3). ProMED was used to verify cases in the media and check for discrepancies. The duration was from 19 March 2014 (first RFI) to 15 October 2014 (index case responsible for infecting two nurses in USA). A total of 31 weeks' ProMED reports were analyzed. Out of 272 ProMED reviewed Only 240 reports were relevant as the others referred to a repeated summary of earlier ProMED reports. Medical evacuations were recorded separately from the import case, due to the assumption that necessary precautions were taken. The first epidemic week is based on the initial RFI report.

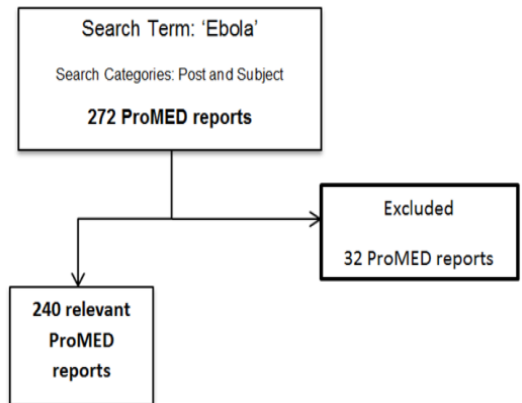

Figure 3: Flow diagram of documents reviewed

Ebola Case Identification-Cases were identified from mass media reports. An initial search was done in "Google News" with keyword "ebola" and date range of 13 Apr 2014 to 18 Sep 2014. Each of the reported patient names was then used to "backtrack" for related cases in Google News. Where patient names were unavailable, either the i) village name and/or ii) hospital name was used. For each case, attempts were made to identify i) the person infecting the patient, ii) others the patient infected, if any, and iii) places the patient traveled to and from while infected, if any. Whenever possible, official sources (*e.g.* World Health Organization) and major news media (*e.g.* CNN) were preferred over unofficial sources and lesser known media; nevertheless, details provided by all retrieved sources were considered for credibility and cogency. Sources for cases tabulated in Table 1, which included.

A. Public News Headline Counts. Two online news databases were queried: ProQuest Newsstand (1,500 newspaper sources) and Dow Jones Factiva (some top newspapers such as AFP and Reuters), using keyword "ebola" for the period of 13 Apr 2014 - 18 Sep 2014.

B. "Google Trends" Search Volumes. Searches were performed in "Google Trends", a public web facility provided by Google that returns the relative search frequencies of search terms and phrases in Google. The keyword "ebola" and a date range of 13 Apr 2014 to 18 Sep 2014 were applied. Weekly data were downloaded for the keyword "ebola" (including search phrases containing it), during the period 13 Apr 2014 - 18 Sep 2014, for the five countries: Guinea, Liberia, Sierra Leone, United States and United Kingdom.

C. Professional Medical Medium. The ProMED (Program for Monitoring Emerging Diseases) emailing system is an internet-based global reporting system widely used by those working with infectious diseases. We searched in ProMED for notifications containing the keyword "ebola", with date ranging from 13 Apr 2014 to 18 Sep 2014.

The analysis of the media attention in Factiva is presented in Table 1. The focus only started after June and picked up from July. The peak of the attention was from August to October. By October, Ebola had spread to USA through an imported case. Medical evacuations of foreign nationals and a Senegalese epidemiologist had been undertaken to USA, Spain, Germany, Norway, France and the United Kingdom (UK).

Table 1: Media focus on Ebola (CNN and BBC)

| Database | Media | Key Word | Time Period | Non Duplicate Search Result |
|---|---|---|---|---|
| Factiva | CNN-All Sources | Ebola | 16 Apr 2014 - 15 Oct 2014 | 1593 |
| Factiva | CNN-All Sources | Ebola | 16 Apr 2014 - 30 Apr 2014 | 11 |
| Factiva | CNN-All Sources | Ebola | 1 May 2014 - 31 May 2014 | 3 |
| Factiva | CNN-All Sources | Ebola | 1 Jun 2014 - 30 Jun 2014 | 4 |
| Factiva | CNN-All Sources | Ebola | 1 July 2014 - 31 July 2014 | 123 |
| Factiva | CNN-All Sources | Ebola | 1 Aug 2014 - 31 Aug 2014 | 412 |
| Factiva | CNN-All Sources | Ebola | 1 Sep 2014 - 30 Sep 2014 | 316 |
| Factiva | CNN-All Sources | Ebola | 1 Oct 2014 - 15 Oct 2014 | 735 |
| Factiva | BBC-All Sources | Ebola | 16 Apr 2014 - 15 Oct 2014 | 650 |
| Factiva | BBC-All Sources | Ebola | 16 Apr 2014 - 30 Apr 2014 | 3 |
| Factiva | BBC-All Sources | Ebola | 1 May 2014 - 31 May 2014 | 1 |
| Factiva | BBC-All Sources | Ebola | 1 Jun 2014 - 30 Jun 2014 | 6 |
| Factiva | BBC-All Sources | Ebola | 1 July 2014 - 31 July 2014 | 40 |
| Factiva | BBC-All Sources | Ebola | 1 Aug 2014 - 31 Aug 2014 | 246 |
| Factiva | BBC-All Sources | Ebola | 1 Sep 2014 - 30 Sep 2014 | 174 |
| Factiva | BBC-All Sources | Ebola | 1 Oct 2014 - 15 Oct 2014 | 181 |

**Results**

One of the objectives of the study is to demonstrate that the transmission dynamic the occurrence of Ebola epidemic can be understood through investigation into the online mass media and social media, including Internet news, web blogs and online search tools such as Google News and Google Trends. We demonstrate that standard hub-and-spoke network diagram of disease transmission reported in the results section can be constructed using cases identified from the online media. Successful construction of these diagrams pivots on the availability and accuracy of the reported relationships (such as kinship or profession) among infected individuals and locations. Most of the time this can be affirmed from observing whether different news sources are reporting consistent information, and from assessing the credibility and immediacy of the sources. Since online news are reported and updated quickly, the dates for case identification are mostly precise.

An initial search was done in "Google News" with keyword "Ebola" and date range of 13 Apr 2014 to 18 Sep 2014. Each of the reported patient names was then used to "backtrack" for related cases in Google News. Where patient names were unavailable, either the i) village name and/or ii) hospital name was used. For each case, attempts were made to identify i) the person infecting the patient, ii) others the patient infected, if any, and iii) places the patient traveled to and from while infected, if any. Whenever possible, official sources (*e.g.* World Health Organization) and major news media (*e.g.* CNN) were preferred over unofficial sources and lesser known media; nevertheless, details provided by all retrieved sources were considered for credibility and cogency (for the sources of cases see Table 2).

The network diagrams of Ebola transmission dynamic synthesized from online media sources can be highly informative in illustrating transmission patterns and suggesting various important controlling factors, such as i) travel, ii) family and village membership, and iii) hospital infections including amongst health care workers. Travel across the borders played a primary role in the spread of Ebola across each of the four West African countries (Figure 4). Traditional treatment seeking across the West African porous borders (Figure 4b and 4c) family visits were one of the many reasons for travel. Most of the reported cases in West Africa involved family and village membership. In Figure 5, the travel of Ebola from Liberia to Nigeria shows how fast the disease can spread via a 'super spreader' was involved in the initial Nigerian outbreak. A total of 20 individuals were infected, 11 were health care workers, and 8 died. Clustering is clearly visible. Figure 6 shows the first imported case, excluding medical evacuations, of Ebola into the US came from Monrovia, Liberia. He continued to infect two nurses at a Dallas hospital despite protective equipment and control measures in place.

Circles at both ends of the travel arrow denote the same person. The reasons for travel were: a) unknown, b) traditional treatment seeking, c) traditional treatment seeking (Sierra Leone to Guinea) and family visit (Guinea to Senegal), d) adoption of Ebola orphan by grandmother. The cases traveling to Senegal and Mali were successfully quarantined.

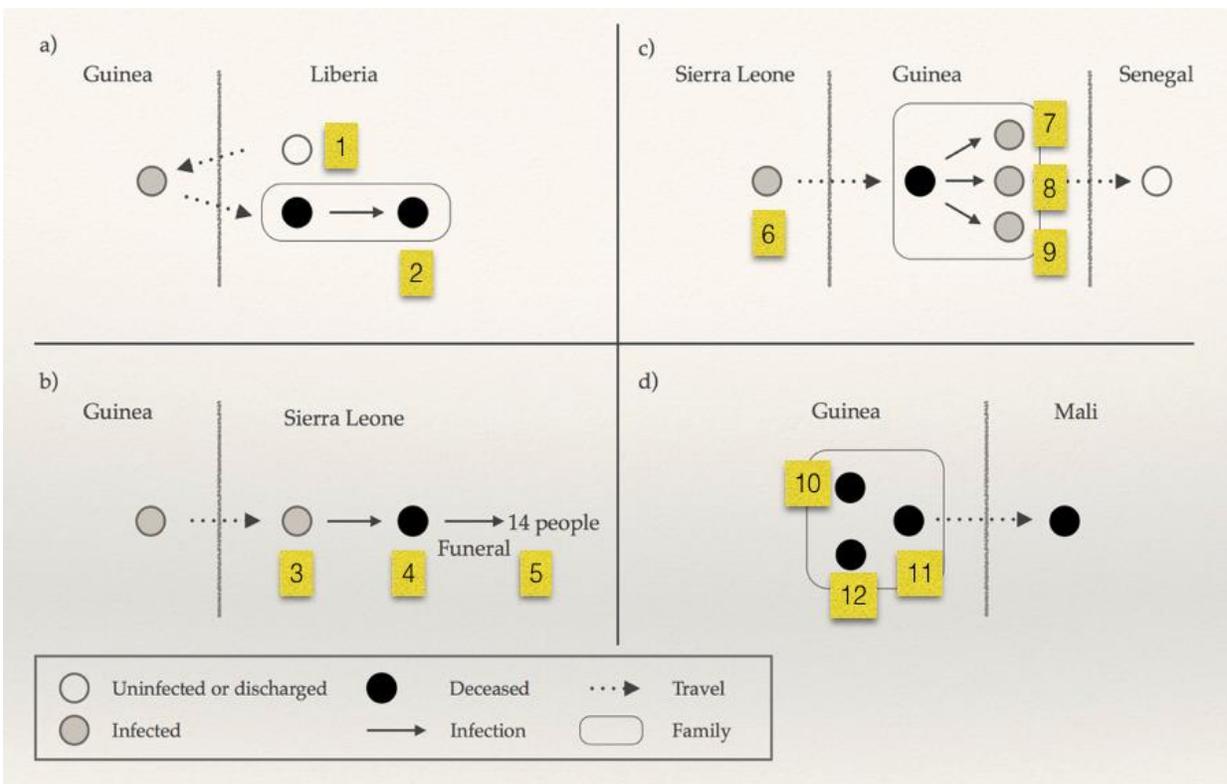

Figure 4. Transmission dynamics of the first reported Ebola cases across each of four West African country borders (a-d)

In Nigeria, a total of 20 people were infected, including 8 who died. Clustering is clearly visible in the diagram, and a "super-spreader" case is involved. Note that circles at both ends of a travel arrow denote the same person. One individual was evacuated from Liberia to Spain for treatment.

Note that circles at both ends of a travel arrow denote the same person. Two nurses at Dallas were infected despite protective equipment and measures. Meanwhile, there were another 6 cases evacuated from West Africa into the US: 4 from Liberia, 1 from Guinea and 1 from Sierra Leone,

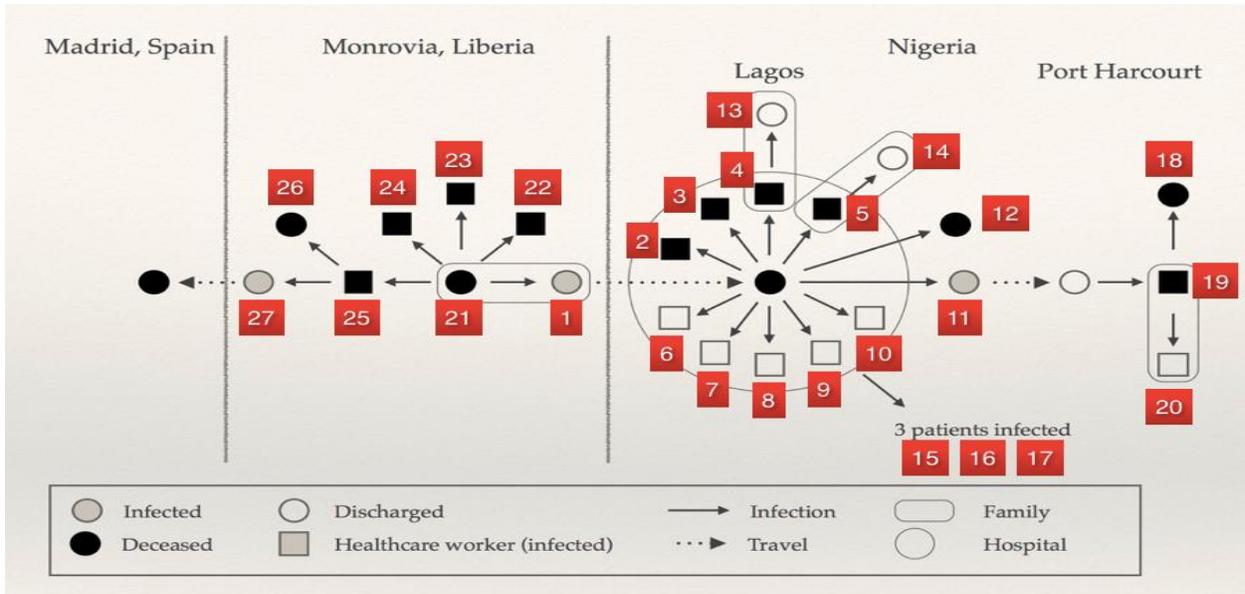

which were all successfully quarantined and treated.

Figure 5. Ebola spread from Liberia to Nigeria during 2014 outbreak

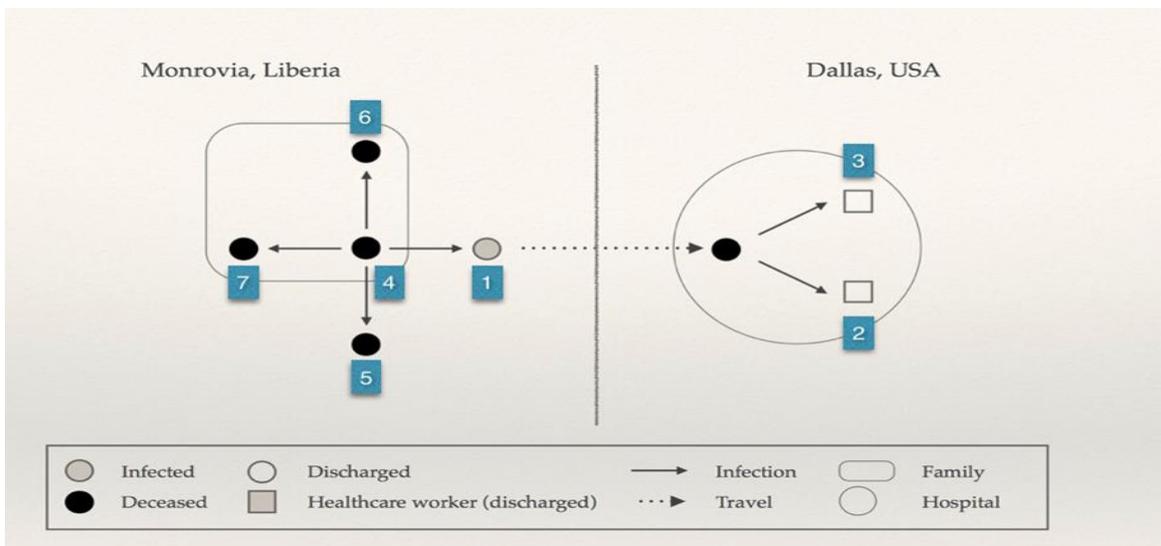

Figure 6. Spread of Ebola from Liberia to the US during 2014 outbreak

Table 2. Sources of Ebola cases identified in Figures 4-6

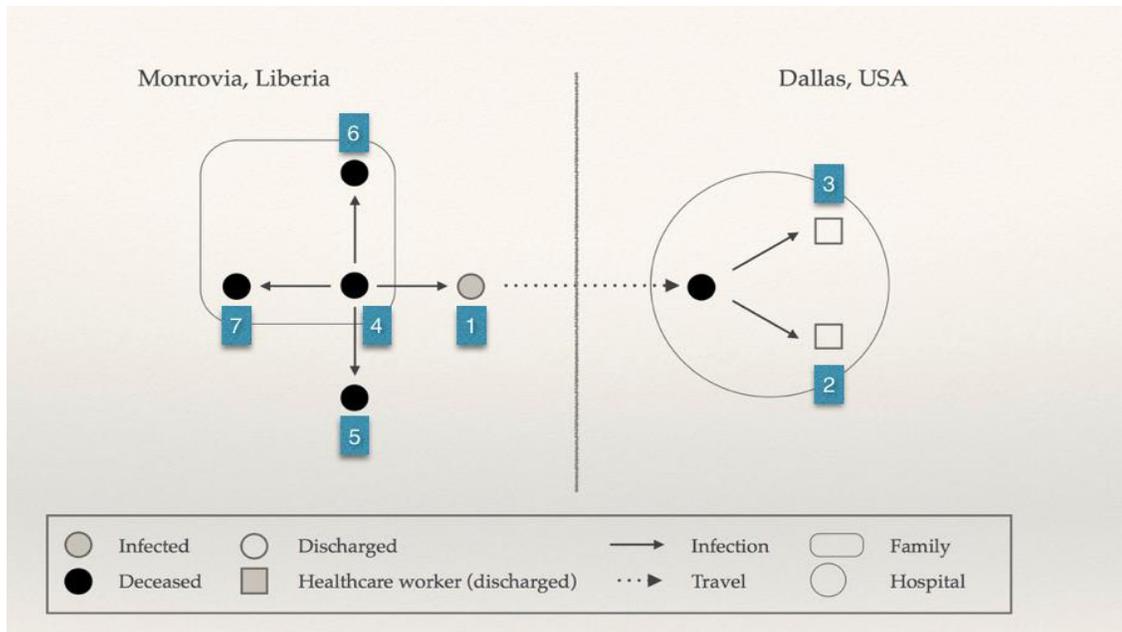

| Diagram | Date Posted | Case(s) | Medium | Title | Link |
|---|---|---|---|---|---|
| **West Africa** | 5/4/2014 | 1,2 | Daily Observer (Liberia) | Ebola claims another victim | http://www.liberianobserver.com/health/ebola-claims-another-victim |
| | 23/9/2014 | 3-5 | WHO | Global Alert and Response (GAR). Sierra Leone: a traditional healer and a funeral | http://www.who.int/csr/disease/ebola/ebola-6-months/sierra-leone/en/ |
| | 9/9/2014 | 6-9 | Reuters | Senegal tracks route of Guinea student in race to stop Ebola | http://www.reuters.com/article/2014/09/09/us-health-ebola-senegal-idUSKBN0H414F20140909/ |
| | 24/10/2014 | 10,11 | CNN | Reports: first confirmed ebola patient in Mali dies | http://edition.cnn.com/2014/10/24/world/africa/mali-ebola/ |
| | 10/11/2014 | 12 | WHO | Mali case, Ebola imported from Guinea | http://www.who.int/mediacentre/news/ebola/10-november-2014-mali/en/ |
| **Nigeria** | 1/10/2014 | 1-20 | WHO | Situation report - 1 October 2014 | http://www.who.int/csr/disease/ebola/situation-reports/archive/en/ |
| | 3/10/2014 | 1-19 | CDC | Morbidity and Mortality Weekly Report. Ebola virus disease outbreak - Nigeria, July-September 2014 | http://www.cdc.gov/mmwr/preview/mmwrhtml/mm6339a5.htm/ |
| | 11/8/2014 | 1-10 | The Guardian | Ebola: Nigeria confirms new case in Lagos | http://www.theguardian.com/society/2014/aug/11/ebola-virus-nigeria-lagos-patrick-sawyer/ |
| | 22/8/2014 | 13,14 | CBS | Ebola makes worrying advance in Nigeria | http://www.cbsnews.com/news/ebola-spreads-in-nigeria-2-new-cases-unconnected-to-patrick-sawyer/ |
| | 1/9/2014 | 18 | Reuters | UPDATE 1-Nigeria records another Ebola case in oil city, 17 cases total | http://www.reuters.com/article/2014/09/01/health-ebola-nigeria-idUSL5N0R24QC20140901/ |
| | 13/8/2014 | 21-27 | Front Page Africa | At Catholic hospital, several ebola deaths traced to Sawyer | http://www.frontpageafricaonline.com/index.php/news/2652-at-catholic-hospital-several-ebola-deaths-traced-to-sawyer/ |
| | 12/8/2014 | 25-27 | The Guardian | Ebola: Spanish missionary dies of disease after being flown to Madrid | http://www.theguardian.com/world/2014/aug/12/ebola-spanish-missionary-dies-madrid-liberia/ |
| **US** | 15/10/2014 | 1-3 | The Daily Mail (UK) | Second ebola-striken nurse, 29, arrives in Atlanta for treatment as it's revealed she called CDC with a fever before boarding a commercial flight to Dallas and they told her she was safe to fly | http://www.dailymail.co.uk/news/article-2793691/second-healthcare-worker-tests-positive-ebola-texas.html/ |
| | 6/10/2014 | 4-7 | CNN | Back in Liberia, ebola is killing Thomas Duncan's neighbors | http://edition.cnn.com/2014/10/06/health/ebola-liberia/ |

From the supplementary data from ProMED, we are aware of a request for information (RFI) on an undiagnosed viral haemorrhagic fever in Guinea on 19 March 2014 based on a report by Standard Media Kenya [45]. Most victims were reported to be in contact with the deceased victim or had handled bodies. By then, at least 35 cases were recorded by local health officials. Dr Sakoba Keita, the Head of the Disease Prevention Unit in Guinea, stated that the symptoms appeared as diarrhoea and vomiting with a very high fever [46]. Some were showing relatively heavy bleeding. The initial suspect disease was Lassa fever or a form of Cholera but Ebola was not excluded from suspicion despite Guinea's position in the Ebola belt of Africa [45, 46]. The confirmation only came on 21 March from a laboratory in Lyons, France [46].

Travel across the borders played a primary role in the spread of Ebola across each of the four West African countries (Figure 4). Traditional treatment seeking across the West African porous borders (Figure 4b and 4c) family visits were one of the many reasons for travel. In **Error! Reference source** **not found.**, the travel of Ebola from Liberia to Nigeria shows how fast the disease can spread via a 'super spreader' was involved in the initial Nigerian outbreak. A total of 20 individuals were infected including 8 who died. Clustering is more visible in this figure. The first imported case, excluding medical evacuations, of Ebola into the US came from Monrovia, Liberia. He continued to infect two nurses at a Dallas hospital despite protective equipment and control measures in place (Figure 6).

The timeline of ProMED headlines, indicating the countries affected, are shown in Figure 7, starting from week 1 of first report from Guinea. The key response events of note reported in ProMED is listed in Figure 8. There are smaller response events (actions by local health authorities) which were not included in the timeline due to repetition of actions. From Figure 8, a visible lag in international response to the outbreak in West Africa can be seen. Of note in Figure 8, there was a visible lag in response to the outbreak in West Africa and by the WHO. Within week 2 of the first reported outbreak, the outbreak had crossed the border to Liberia and then onto Sierra Leone. Despite the World Health's Organization's (WHO) recommendation not to shut the borders, the neighboring African nations started to close the borders with almost little success because of the porous borders and trade between countries. Within week 19 of the epidemic, Nigeria was affected. In week 28, the first EVD patient in USA walked into a hospital and was sent home. By week 30, a healthcare worker in US who cared for him was tested positive for the disease. The WHO only declared it an emergency in week 21 (August) and started situational updates on 15 September 2014.

Figure 7: ProMED headlines indicating location of confirmed/suspected/probable cases from the first RFI

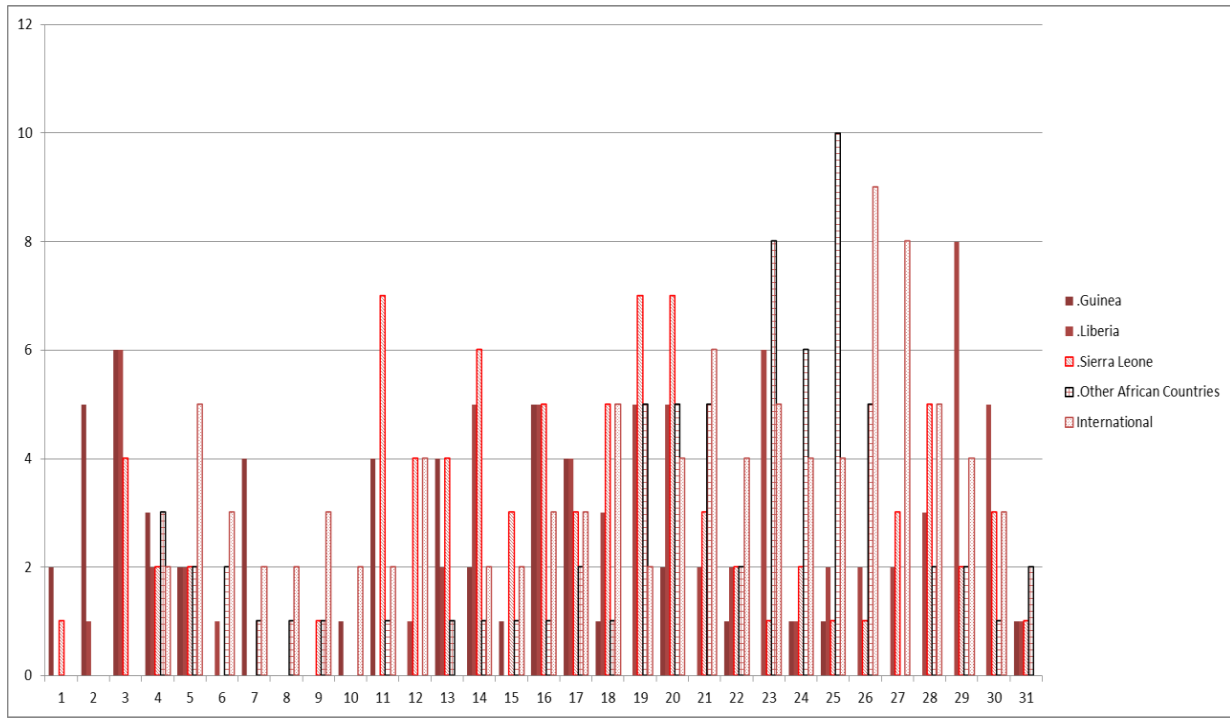

Figure 8: Selected initial key responses to the West African Ebola epidemic

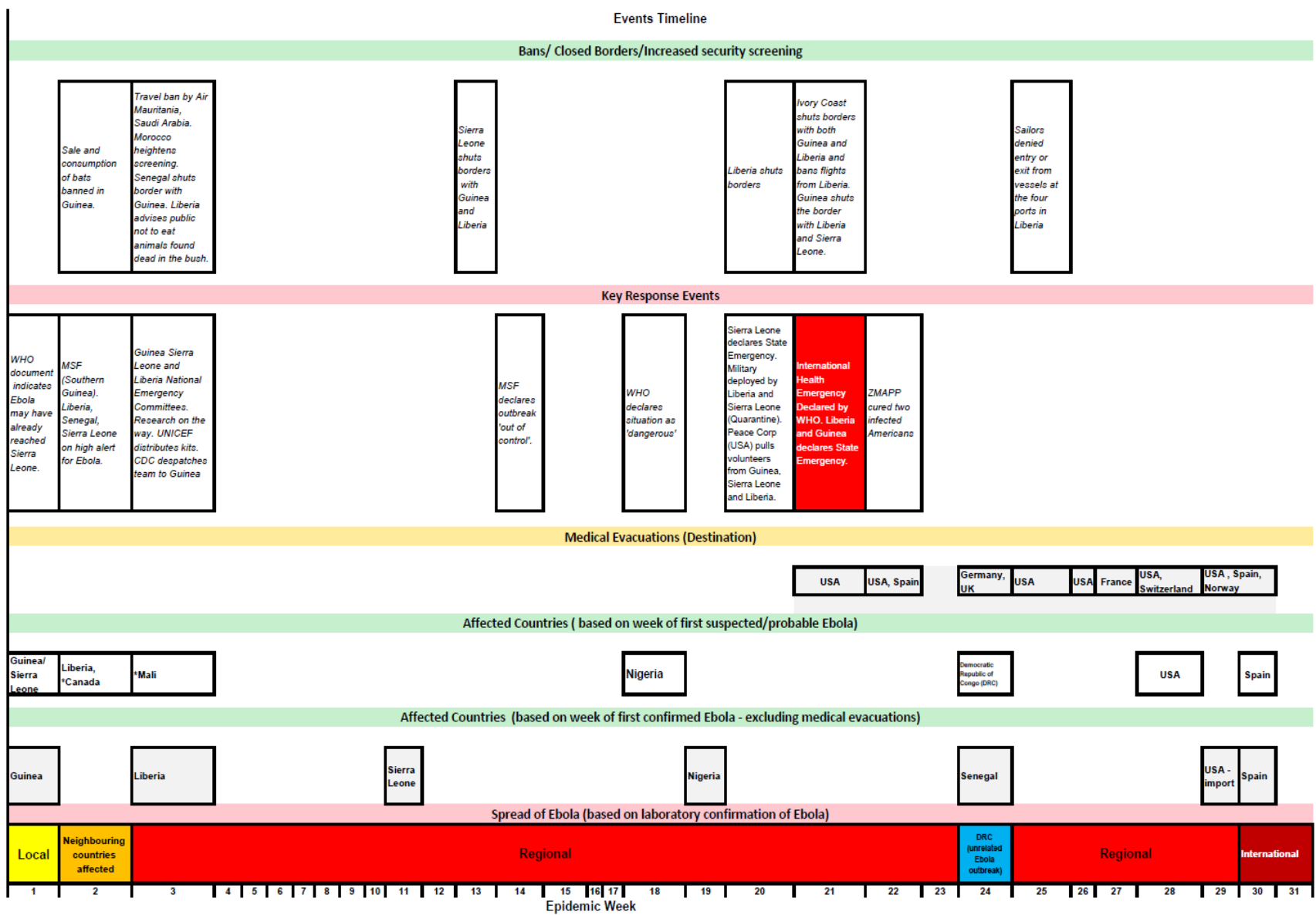

**Conclusion**

There are preliminary findings that the disease could have started earlier in December 2013[33]. Based on both media and ProMED reports, the first sign of the problem was signaled by ProMED in early March as an undiagnosed viral haemorrhagic disease. The initial responders were Médicins San Frontières (MSF), Centers for Disease Control (CDC), and the World Health Organization (WHO). There were visible time gaps between the international responses to Ebola based on the ProMED timeline despite MSF public declaration that Ebola was out of control in West Africa. By July, the international media interest started to increase on Ebola. In comparing the responses by two news agencies, CNN, an American based news network, had more interest in Ebola than the BBC because amongst the first countries to medically evacuate their nationals back was USA while the UK had not started medical evacuations yet. This suggests that the public is more interested in Ebola if it enters their home country, rather than an outbreak happening in another continent. Due to the notion of interconnected and interdependencies of the world, we need to view the world as a single interconnected system where localized problems, if not contained well, could propagate fast to more locations affecting the globe. This has significant implications for improving the detection, preparedness and response of future disease outbreaks for the disaster medicine and public health preparedness community.

It can be readily seen how digital surveillance network becomes a most useful and powerful tool when tracking the diffusion of Ebola in a neighborhood, community or region. In a developing country setting, like West Africa, digital surveillance may not be effective for early disease detection, due to the lack of access to the communication (e.g., social media) and healthcare resources [3]. Instead, the digital surveillance could be used as an early response tool to curb an outbreak, like Ebola, which also have implications for the current COVID-19 as well as preparedness and response of future outbreaks. It is evident that flow of information and robustness to support the flow of information was instrumental in not recognizing or early detection to this local problem, which is now affecting the entire globe. Development of community partnership and trust through digital networks could cultivate shared responsibility in dealing with administrative, cross cultural and other socio economic barriers for building better information networks for supporting future disaster medicine and public health preparedness efforts.